# Quantum Landau–Lifshitz–Gilbert equation in ferromagnetic conductors and magnetic insulators

Ji-Yang Zhu and Zheng-Chuan Wang*
*School of Physical Sciences, University of Chinese Academy of Sciences, Beijing 100049, China*

Understanding the microscopic origin of magnetic damping is essential for describing magnetization dynamics in nanoscale magnetic systems. In this work, we develop a microscopic framework to derive Gilbert damping terms from specific interactions in ferromagnetic conductors and magnetic insulators. In ferromagnetic conductors, magnetic damping is dominated by the s–d exchange interaction between the localized magnetic moments and the itinerant electrons, which is effectively equivalent to an electron–magnon interaction. From the spin diffusion equation and the spin-transfer torque therein, we obtain the corresponding quantum LLG equation. In magnetic insulators, where free electrons are absent, spin–phonon interactions become the primary source of the damping.Starting from an effective spin–phonon interaction and using linear response theory, we derive a quantum LLG equation with a Gilbert damping tensor analytically. Our results demonstrate that different microscopic interactions lead to qualitatively different structures of magnetic damping. Instead of being described by a universal scalar parameter, Gilbert damping can exhibit spatial dependence and tensorial characteristics determined by the underlying microscopic interactions.

A central issue in magnetism is understanding the microscopic origin of magnetization relaxation. The Landau–Lifshitz–Gilbert(LLG) equation [1], which successfully describes the relaxation of magnetic moments toward equilibrium [2] has been widely applied in various aspects of magnetic research, playing a crucial role in many fields such as domain walls [3], skyrmions [4], and the thermodynamic properties of magnetization dynamics [5]. In the conventional formulation, the Gilbert damping coefficient is usually introduced phenomenologically as a scalar parameter. However, its microscopic origin depends strongly on the interaction between magnetic degrees of freedom and their surrounding environment.

Experimentally observed Gilbert damping generally contains both intrinsic (spin–phonon, spin–electron coupling) and extrinsic (lattice defects) contributions[6] [7] [8]. Beyond these mechanisms, additional effects, such as relativistic contributions, may also play a role.[9] As the research scale shrinks to the nanoscale, microscopic interactions between magnetic moments and their environment become increasingly important, and a phenomenological description of damping may become insufficient, for example current-induced magnetic switching in single-molecule magnets can be explained by considering the quantum effect [10].

Recent studies have developed microscopic theories of Gilbert damping by explicitly considering the coupling between magnetization and its surrounding environment. In ferromagnetic conductors, electron-mediated damping has been shown to exhibit nonlocal and anisotropic features beyond the conventional scalar Gilbert damping [11] [12]. In magnetic insulators, spin-phonon interactions and phonon-mediated effects have also been identified as important mechanisms affecting magnetization dynamics. [13] [14] These works demonstrate that the structure of Gilbert damping is closely related to the microscopic nature of the coupled bath, and different interactions may lead to distinct damping behaviors.

In this work, we investigate how different microscopic interactions generate Gilbert damping terms in ferromagnetic conductors and magnetic insulators. We further demonstrate how different microscopic mechanisms lead to distinct forms of damping within a unified framework: in ferromagnetic conductors, the damping is governed by spin transport processes and becomes explicitly dependent on spatial variations of the magnetization, while in magnetic insulators, it is determined by spin–phonon interaction and gives rise to a tensorial damping structure.

*Ferromagnetic conductor*—In ferromagnetic conductors, magnetic damping can originate from several microscopic mechanisms, including electron-mediated processes and spin-phonon interactions. In ferromagnetic conductors, the contribution from itinerant electrons is generally considered to be one of the dominant damping channels, while spin-phonon effects are typically weaker. In this work, we focus on the damping induced by the s-d exchange interaction between itinerant electrons and localized magnetic moments. The interaction between itinerant electrons and the magnetization $\vec{M}(|\vec{M}| = 1)$ is usually described by the s-d exchange model. When the localized spins are expressed in terms of magnon operators, this interaction can lead to an effective electron–magnon coupling, providing a microscopic origin of electron-mediated magnetic damping. [15] The Hamiltonian of the itinerant electron can be written as

$$\hat{H} = \frac{\vec{p}^2}{2m_e} - J_{ex}\hat{\vec{\sigma}} \cdot \vec{M}(\vec{r}, t), \tag{1}$$

where $\hat{\vec{\sigma}}$ is the Pauli spin matrices, $J_{ex}$ is the s-d exchange constant.

As shown in Ref. [16], if magnetization varies with space and time slowly, it can be treated as an effective electro-magnetic field perturbation for itinerant electrons and can induce a spin current in the conductor. This provides a connection between magnetic dynamics and spin transport. The effective electromagnetic field corresponds to the magnetic texture in ferromagnet is

$$E_i = \pm (\frac{\hbar}{2e})(\partial_t \vec{M} \times \partial_i \vec{M}) \cdot \vec{M}, \quad (2)$$

$$B_i = \mp (\frac{\hbar}{2e})\epsilon^{ijk}(\partial_j \vec{M} \times \partial_k \vec{M}) \cdot \vec{M}, \quad (3)$$

where $\pm$ stands for the spin up and down bands, $i, j, k$ mean the $x, y, z$ component, $\epsilon^{ijk}$ is the Levi-Civita symbol and the Einstein summation convention is implied. In most transition metal ferromagnet, the dependence of conductivity on the effective magnetic field (magnetoresistance effect) is weak and is therefore neglected, and the spin current driven by the variation of magnetization is given by

$$j_s^i = -\frac{g\mu_B}{2e}(G^+ E_i^+ - G^- E_i^-)\vec{M} = -\frac{g\mu_B \hbar G_0}{4e^2}[(\partial_t \vec{M} \times \partial_i \vec{M}) \cdot \vec{M}]\vec{M}, \quad (4)$$

which expresses the spin current in the $i$ direction, where $G^{\pm}$ is the conductivity for the spin up and spin down bands, $G_0 = G^+ + G^-$, $\mu_B$ is the Bohr magnon and $g$ is the Land´e g-factor.

* Contact author: wangzc@ucas.ac.cn

Considering the above spin current induced by the variations of the magnetization, we can obtain the spin accumulation $\vec{m}(\vec{r}, t)$ as [17]

$$\vec{m}(\vec{r}, t) = -\frac{\tau_{sf}}{1+\xi^2}((\vec{\nabla}_r \cdot \vec{\vec{j}}_s) + \xi \vec{M} \times (\vec{\nabla}_r \cdot \vec{\vec{j}}_s)) + <\vec{m}>, \quad (5)$$

where $\xi = \tau_{sf}\frac{J_{ex}}{\hbar}$, $\tau_{sf}$ is the spin relaxation time, $\vec{\vec{j}}_s$ is the spin current density which is a tensor. Substituting Eq. (4) into Eq. (5) and expanding the spin accumulation around the adiabatic accumulation $<\vec{m}>$, we can obtain the non-adiabatic deviation $\delta\vec{m}$ from $<\vec{m}>$. Since the induced spin current and spin deviation are small, we ignore the feedback of the deviation on the spin current and the higher order derivative of $\vec{m}$, the resulting deviation of the spin accumulation $\delta\vec{m}$ is

$$\delta\vec{m} = \frac{\tau_{sf}}{1+\xi^2}\frac{g\mu_B \hbar G_0}{4e^2}[[(\partial_i \vec{M} \times \vec{M}) \cdot \partial_t \vec{M}]\partial_i \vec{M} + \xi \vec{M} \times \partial_i \vec{M}[(\partial_i \vec{M} \times \vec{M}) \cdot \partial_t \vec{M}]]. \quad (6)$$

For magnetization dynamics, the adiabatic part of the spin accumulation remains parallel to the local magnetization and therefore does not contribute directly to the damping progress. As a result, the damping originates from the nonadiabatic spin accumulation, which contributes through the Levy spin torque, $\frac{J_{ex}}{\hbar}\vec{M} \times \vec{m}$ indicating that $\vec{m}$ can be regarded as an effective mag -netic field acting on the magnetization. Substituting these results into the Heisenberg equation of motion for the spin operator which is the microscopic source of the local magnetization, we obtain

$$\frac{\partial \hat{\vec{S}}}{\partial t} = -\gamma \hat{\vec{S}} \times \vec{H}_{eff} + \frac{\xi}{1+\xi^2}\frac{g\mu_B \hbar G_0}{4e^2}[\hat{\vec{S}} \times \partial_i \vec{M}[(\vec{M} \times \partial_i \vec{M}) \cdot \partial_t \vec{M}] + \xi \hat{\vec{S}} \times (\vec{M} \times \partial_i \vec{M})[(\vec{M} \times \partial_i \vec{M}) \cdot \partial_t \vec{M}]], \quad (7)$$

where $\gamma$ is the gyromagnetic ratio, and magnetization $\vec{M}$ can be given by the expectation of the spin operator $\vec{M} = \sum_j \frac{g\mu_B}{V} < \hat{\vec{S}}_j >$, where $V$ is a volume element. $\vec{H}_{eff}$ is the effective field which usually contains exchange field, Dzyaloshinskii-Moriya interaction and so on. This equation leads to two Gilbert-like damping terms whose components are given by

$$\alpha_{1,mn} = \frac{\xi}{1+\xi^2}\frac{g\mu_B \hbar G_0}{4e^2 M_s}(\partial_i \vec{M})_m (\vec{M} \times \partial_i \vec{M})_n,$$

$$\alpha_{2,mn} = \frac{\xi^2}{1+\xi^2}\frac{g\mu_B \hbar G_0}{4e^2 M_s}(\vec{M} \times \partial_i \vec{M})_m (\vec{M} \times \partial_i \vec{M})_n, \quad (8)$$

where the subscripts $m$ and $n$ refer to the spatial compo -nents $x, y, z$, $M_s$ is the saturation magnetization. Unlike the conventional scalar Gilbert damping, the damping terms obtained here contain two distinct tensor contributions $\alpha_1$ and $\alpha_2$. These terms originate from the misalignment between the spin accumulation and the local magnetization direction, which generates a dissipative spin-transfer torque. As a result, the magnetic relaxation depends explicitly on the spatial variation of the magnetization and requires a tensorial description of the damping coefficient. The spatially dependent damping obtained here is closely related to the nonlocal damping discussed previously in ferromagnetic conductors. In particular, Ref. [16] showed that dynamic magnetic textures can generate spin currents and dissipative spin-transfer torques through spin accumulation. Our result provides a complementary description by explicitly relating the damping tensor to the spin accumulation induced by the s-d exchange interaction, and provides a microscopic connection between electron-mediated interactions and the resulting damping structure.

According to the analytical expression derived above, these two terms relate the magnetization damping to its spatial variations, and the corresponding damping tensor $\alpha_1$ is non-symmetric, indicating anisotropic and nonlocal magnetic damping. We further provide an order-of-magnitude estimate of the results using parameters usually found in ferromagnetic conductors, the exchange energy is adopted as $J_{ex} \sim 1eV$, and the spin–flip relaxation time is $\tau_{sf} \sim 10^{-12}s$, which leads to the parameter $\xi \sim 10^3$, $G_0 = 1.5 \times 10^7 s.m^{-1}$ and $M_s = 8 \times 10^5 A \cdot m^{-1}$. For a magnetization texture varying over a characteristic wavelength of approximately 30nm, we estimate the symmetric damping coefficient to be $\alpha_{2,\alpha\beta} \sim 0.15$. This estimated Gilbert damping coefficient is of the same order of magnitude as that measured experimentally in ferromagnetic conductors [18] [19] [20]. The asymmetric contribution is much smaller than the symmetric contribution $\alpha_2$, with $\frac{\alpha_{1,\alpha\beta}}{\alpha_{2,\alpha\beta}} \approx 10^{-3}$ rendering the former negligible in practice. But for some special cases, whose exchange energy and spin–flip relaxation time are a bit smaller than normal, $\xi$ can be reduced to around 10, in this case the asymmetric damping can not be ignored and may have a noticeable effect on the motion of magnetization.

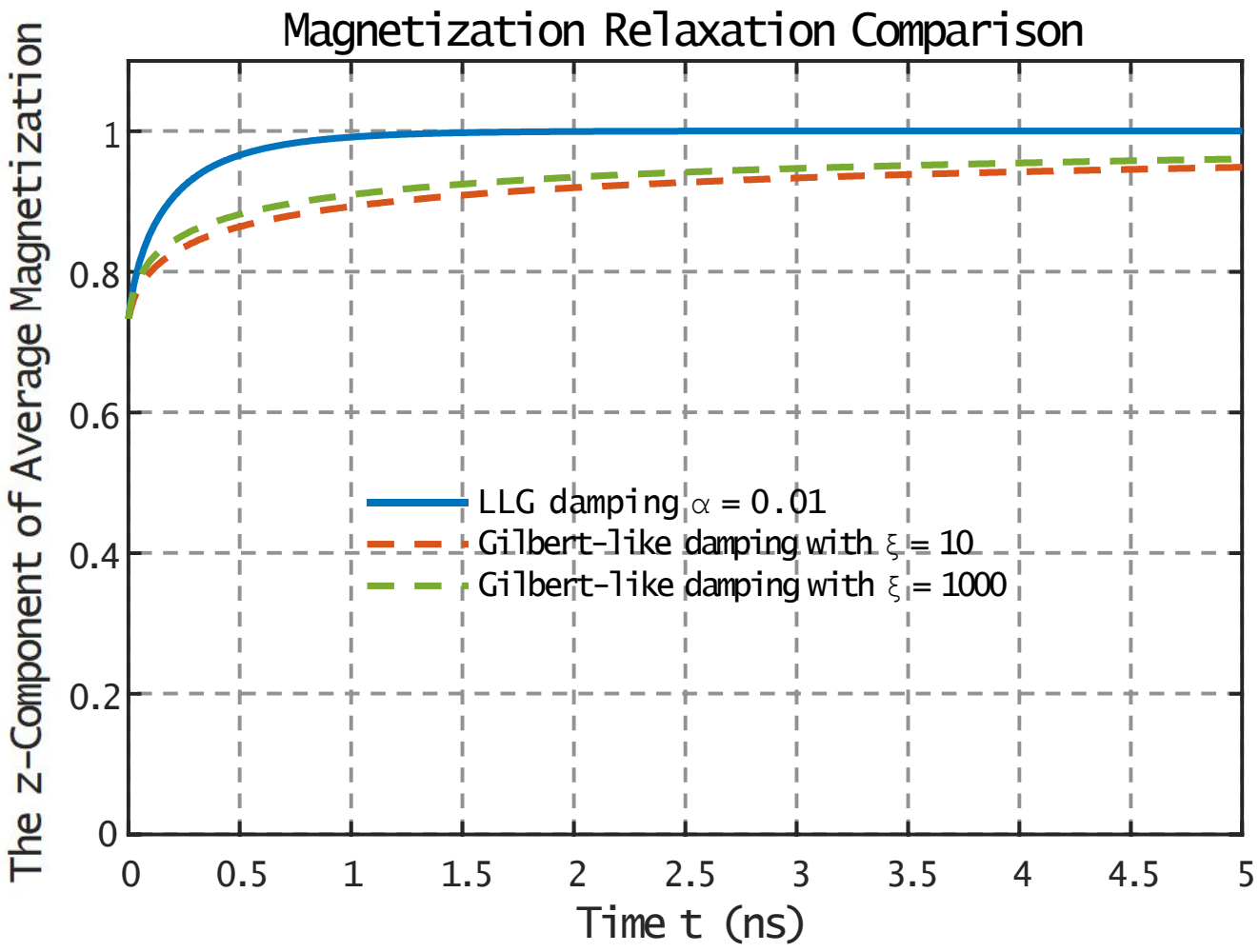

Figure 1. Time evolution of the $z$-component of the average magnetization for three different types of Gilbert damping: classical LLG damping with $\alpha = 0.01$, our nonlocal damping with $\xi = 10$, and our nonlocal damping with $\xi = 1000$.

In the numerical simulation, we consider a two-dimensional ferromagnet with a size of 80nm × 80nm and compare the magnetization dynamics governed by the quantum LLG equation with that described by the classical LLG equation. The effective field includes only the external magnetic field and the exchange coupling field. Using the parameters specified above, as shown in Fig.1 we find a clear difference between the magnetization dynamics predicted by the quantum and classical LLG equations, for the z-component, the quantum case exhibits a rapid initial reduction of magnetization, followed by a much slower relaxation toward the equilibrium state with magnetization aligned along the z-axis. We also compare the effects of different values of $\xi$ on the magnetic damp -ing. At the very beginning of the evolution, the difference between the two cases is negligible. As the damping proceeds, however, the contribution of the asymmetric term becomes more evident for $\xi = 10$, leading to a slower relaxation toward the equilibrium state.

*Magnetic insulator*—For magnetic insulators, in the absence of itinerant electrons, spin-phonon interactions provide an important mechanism contributing to magnetic relaxation. Recent studies have demonstrated that coupling between spin dynamics and phonon degrees of freedom can generate intrinsic damping and modify magnetization dynamics through phonon-mediated effects[13]. In this work, we investigate how spin-phonon interactions give rise to the tensorial structure of Gilbert damping.

Motivated by the phonon induced modification of intersite hopping, we employ a two-orbital Hubbard model with spin-orbit coupling to describe the electron–phonon interaction [21]. Using the Brillouin-Wigner perturbation theory, up to the second order, the term in the effective Hamiltonian that leads to a Raman-type spin–phonon interaction [22] can be written as

$$H_{eff} = \sum_{\substack{i,n,m \\ m\neq n}} \frac{2}{\Delta_{12}^2} (\hat{u}_{i,n}\hat{p}_{i,m} - \hat{p}_{i,n}\hat{u}_{i,m})[(\vec{\delta}_{i,n}^{12} \times \vec{\delta}_{i,m}^{21} - d_{i,n}^{12}\vec{\delta}_{i,m}^{21} - d_{i,m}^{21}\vec{\epsilon}_{i,n}^{12}) \cdot \hat{\vec{S}}_i], \quad (9)$$

where $\hat{u}_{i,n}$, $\hat{p}_{i,n}$ denote the displacement from its equilibrium position and momentum of the i-th ion along the n -direction, $\Delta_{12} = \epsilon_2 - \epsilon_1$ is the energy difference between two orbitals on the same site, n, m $\in$ {x, y, z}. $\vec{\delta}_{i,n}^{\alpha\beta}$ and $d_{i,n}^{\alpha\beta}$ are the expansion coefficients of the onsite transition matrix $D_{i,n}^{\alpha\beta} = d_{i,n}^{\alpha\beta}\hat{I} + i\vec{\delta}_{i,n}^{\alpha\beta} \cdot \hat{\vec{\sigma}}$ in the electron-phonon interaction Hamiltonian [21]. In the Hubbard model $U^2 \gg \Delta_{12}^2$ for insulator, where U denotes correlation strength, the effective on-site interaction $H_{eff}$ becomes dominant. For simplification, we assume that this interaction can be simplified to $-g\hat{\mathbf{J}}_i \cdot \hat{\mathbf{S}}_i$. We consider a magnetic insulator in an external magnetic field with the nearest-neighbor Heisenberg exchange interaction. Other interactions are neglected since they do not qualitatively affect our conclusions. Under these assumptions, after taking average over the phonon degrees of freedom, the equation of motion for the spin operator of the i-th ion can be obtained directly from the Heisenberg equation as

$$\frac{d}{dt}\hat{\vec{S}}_i(t) = -\hat{\vec{S}}_i(t) \times [\gamma\vec{H}(t) + \sum_j J_{ij}\hat{\vec{S}}_j(t)] + g\hat{\vec{S}}_i(t) \times < \hat{\vec{J}}_{ph,i}(t) >, \quad (10)$$

where $\mathbf{H}$ is the external magnetic field, $\hat{\mathbf{S}}_i$ is the spin operator of the i-th ion, $J_{ij}$ is the Heisenberg exchange coefficient between i-th and j-th ions, $\hat{\mathbf{J}}_{ph,i}$ is the angu lar momentum of i-th ion, and $< \cdots >$ expresses the statistical average over different phonon states.

Due to the spin–phonon interaction, spin dynamics is coupled to the ionic motion. Therefore, when considering the motion of ion, we must also take into account the effect of the spin, so we decompose the spin-phonon interaction into the static part $H_{s-ph}^{eq} = -g\sum_i \hat{\vec{J}}_i \cdot \hat{\vec{S}}_i^{eq}$ and the deviation from it,$i.e$ the dynamical part $H_{s-ph}^{dyn} = -g\sum_i \hat{\vec{J}}_i \cdot \delta\hat{\vec{S}}_i$ where $\hat{\vec{S}}_i^{eq}$ represents the spin operator for the i-th ion at equilibrium and $\delta\hat{\mathbf{S}}_i$ denotes its deviation from equilibrium, the statistical average of the μ-component of $< J_{i,\mu}(t) >$ for the i-th ion is given by the Kubo formula with respect to the variation $\delta\hat{\mathbf{S}}_i$ as

$$< \hat{J}_{i,\mu}(t) >_{H_{ph}+H^{eq}_{S-ph}(t)} = < \hat{J}_{i,\mu} >_{H_{ph}+H^{eq}_{S-ph}} - \frac{ig}{\hbar}\sum_\nu \int_0^\tau dt' < [\hat{J}_{i,\mu}(t'), \hat{J}_{i,\nu}] >_{H_{ph}+H^{eq}_{S-ph}} (\delta\hat{S}_{i,\nu}(t) - t'\frac{d}{dt}\hat{S}_{i,\nu}(t)), \tag{11}$$

Under the conditions where the dominant phonon frequencies are comparable to or lower than the magnon frequencies and anharmonic scattering ensures fast phonon relaxation, the relaxation time of the phonon bath is much shorter than the characteristic timescale of spin dynamics; hence the correlation time is effectively short. This assumption corresponds to a Markovian approximation, which justifies extending the upper limit of the integral to infinity and neglecting memory effects, so we can obtain a Gilbert damping term as

$$\frac{ig^2}{\hbar}\hat{\vec{S}}_i(t) \times \sum_\nu \int_0^\infty dt' < [\hat{\vec{J}}_i(t'), \hat{J}_{i,\nu}] >_{H_{ph}+H^{eq}_{S-ph}} t'\frac{d}{dt}\hat{S}_{i,\nu}(t)) , \tag{12}$$

in which the linear response function for the $\mu\nu$-components of the phonon angular momentum can be defined as

$$\chi_{\mu\nu}(z) = \frac{1}{i\hbar}\int_0^\infty dt e^{izt} < [\hat{\vec{J}}_i(t), \hat{J}_{i,\nu}] >_{H_{ph}+H^{eq}_{S-ph}}, \tag{13}$$

where $\approx$ is a complex number. However, in materials where the phonon spectrum is dominated by high-frequency modes, the approximation expansion adopted here may not be quantitatively accurate, and additional damping mechanisms or non-Markovian effects may need to be considered. Considering the diagonal components of Gilbert damping tensor as the dominant contribution, if we decompose the complex number $\approx$ into its real part $\Omega$ and imaginary part $\eta$, we can express it as

$$\alpha_{\mu\mu} = -g^2 \lim_{\eta\to 0+}\lim_{\Omega\to 0} Im\frac{\chi_{\mu\mu}(\Omega+i\eta)}{\Omega} = g^2\pi\beta S_{\mu\mu}(0), \tag{14}$$

where $\beta = \frac{1}{k_B T}$, T is the temperature, $S_{\mu\mu}(\omega) = \frac{1}{2\pi}\int_{-\infty}^{\infty} < \hat{\vec{J}}_{i,\mu}(t)\hat{J}_{i,\mu} >_0 e^{i\omega t}dt$ is the dynamical structure factor of $\hat{J}_{i,\mu}$.

From the phonon angular momentum theory proposed by Zhang and Niu[23], the angular momentum of i-th ion is

$$\hat{\vec{J}}_i(t) = \frac{-i\hbar}{2N}\sum_{k,k'}(\epsilon^*_k \times \epsilon_{k'})(\sqrt{\frac{\omega_k}{\omega_{k'}}}a_{k'}a^+_k + \sqrt{\frac{\omega_{k'}}{\omega_k}}a^+_k a_{k'})\, e^{i(\vec{k}'-\vec{k})\cdot\vec{R}_l}e^{i(\omega_k-\omega_{k'})t}, \tag{15}$$

where $k = (\mathbf{k}, \sigma)$ specifies the wave vector and branch, $E_k$ is the displacement polarization vector, $a^\dagger_k$ is the Phonon creation operator for mode k, $\omega_k$ is the phonon frequency of mode k and N is total number of ions. Substituting this expression into $S_{\mu\mu}$, we can obtain

$$S_{\mu\mu}(0) = -\frac{\hbar^2}{4N^2}[\sum_{k_1,k_2}\delta(\omega_{k_1}-\omega_{k_2})(\epsilon^*_{k_1}\times\epsilon_{k_2})_\mu(\epsilon^*_{k_2}\times\epsilon_{k_1})_\mu 4f_{k_1}(f_{k_2}+1) + \sum_{k_1,k_2,k_3,k_4}\delta(\omega_{k_1}-\omega_{k_2})(\epsilon^*_{k_1}\times\epsilon_{k_2})_\mu(\epsilon^*_{k_3}\times\epsilon_{k_4})_\mu(2f_{k_1}+1)(2f_{k_3}+1)\delta_{k_1k_2}\delta_{k_3k_4}], \tag{16}$$

where the superscript * denotes the complex conjugation, $f_i = (e^{\beta\hbar\omega_i}-1)^{-1}$ is the Bose-Einstein distribution function. We consider a cubic lattice with the unperturbed state in which all spins are aligned along the z-axis. In this case the phonon eigenvectors naturally separate into two distinct types, one type lies in the x–y plane and consists of two eigenvectors, while the other is oriented along the z-axis and consists of a single eigenvector. Consequently, the first term yields a tensorial Gilbert damping and the second term contributes only to the zz component of the Gilbert damping tensor. The resulting equation of motion for the spin operator can therefore be written as

$$\frac{\partial\hat{\vec{S}}}{\partial t} = -\gamma\hat{\vec{S}}\times\vec{H}_{eff} + \hat{\vec{S}}\times(\vec{\vec{\alpha}}\cdot\frac{\partial\hat{\vec{S}}}{\partial t}) + \lambda\hat{\vec{S}}\times\vec{z}\frac{\partial\hat{S}_z}{\partial t}, \tag{17}$$

where $\vec{\vec{\alpha}}$ is the Gilbert damping tensor, $\lambda$ quantifies the extra damping induced by the temporal evolution of the z-component of spin. This equation represents the interaction-induced LLG dynamics for magnetic insulators derived from the microscopic spin-phonon interaction. Compared with the conventional scalar Gilbert damping, the present formulation contains a tensorial damping coefficient arising from the correlations between spin dynamics and phonon angular momentum. This tensorial form describes the directional dependence of energy dissipation induced by lattice vibrations. The anisotropic damping obtained here originates from the directional correlation of phonon angular momentum fluctuations. Such anisotropic dissipation is consistent with the general expectation that different microscopic baths can produce different damping structures [13]. In real materials, crystal symmetry and lattice anisotropy may further modify the damping tensor, leading to additional anisotropic contributions.

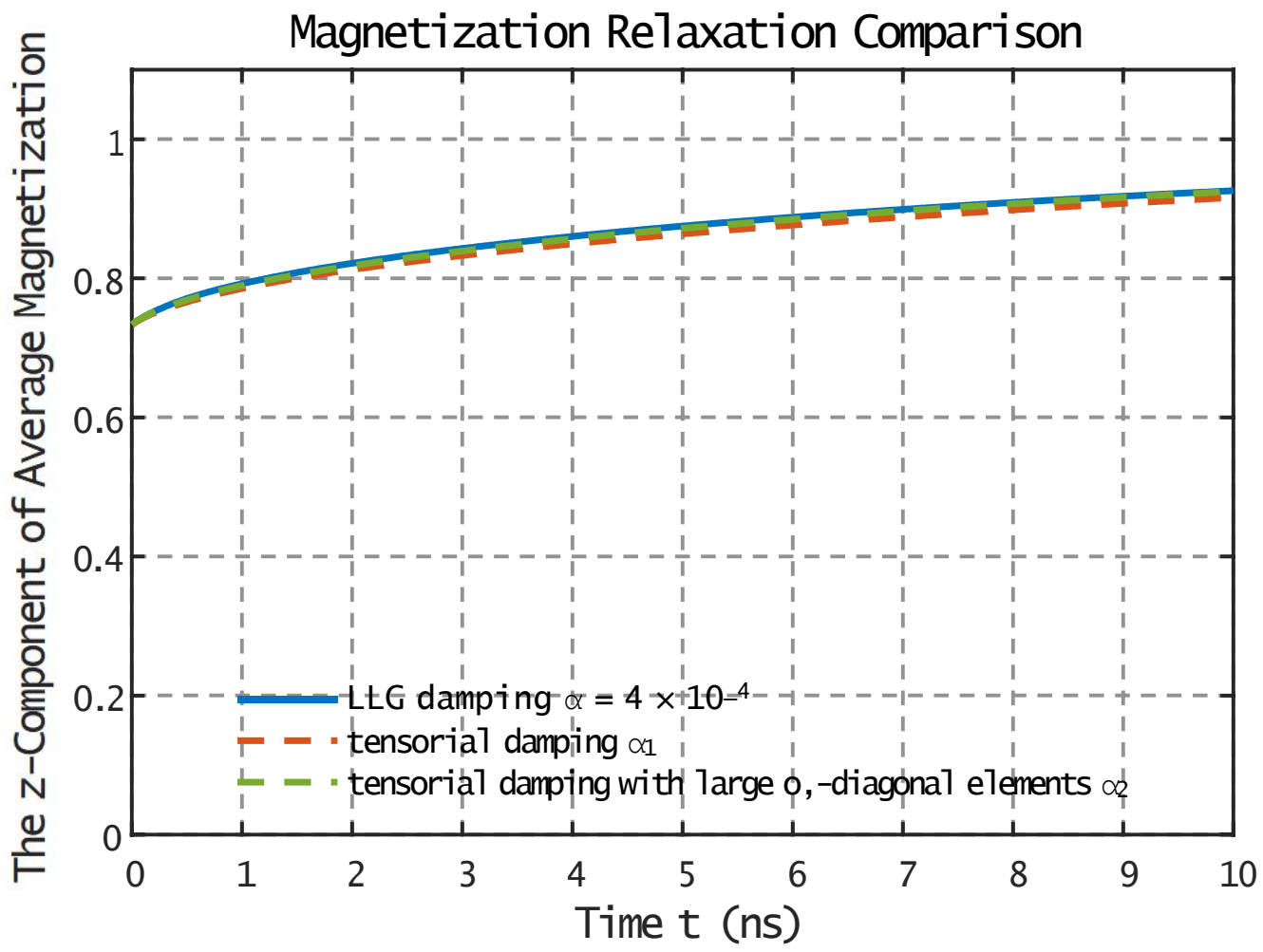

Figure 2. The time evolution of z-component of the average magnetic moment for three different types of Gilbert damping: the classical LLG damping with $\alpha = 4 \times 10^{-4}$, the damping with a tensorial coefficient $\alpha_1$, the damping with a tensorial coefficient $\alpha_2$.

We estimate the contribution of phonons to the Gilbert damping with the following parameters: $T = 10K$, thelongitudinal spring constant is $K_L = 0.144 eV/(u\mathring{A}^2)$ and the transverse spring constant is $K_T = K_L/4$, the unit cell lattice vectors are $(a, 0, 0)$, $(0, a, 0)$, $(0, 0, a)$ with $a = 1\mathring{A}$ as given by [23]. For an effective spin–phonon coupling strength of $\lambda \approx 4THz$, the diagonal components of the Gilbert damping tensor are found to be of order $10^{-4}$, which is comparable to the experimentally reported values for several magnetic insulators at low temperatures [24] [25].

Using the same initial magnetization configuration mentioned in the ferromagnetic conductor case, we numerically compare the magnetic damping described by the quantum LLG equation with that obtained from the classical LLG equation. Since the $\lambda$ term can be absorbed into the Gilbert damping tensor as an additional zz-component, we choose two different damp - ing tensor, $\vec{\vec{\alpha}}_1 = \begin{pmatrix} 4 \times 10^{-4} & 2 \times 10^{-5} & 2 \times 10^{-5} \\ 2 \times 10^{-5} & 4 \times 10^{-4} & 2 \times 10^{-5} \\ 2 \times 10^{-5} & 2 \times 10^{-5} & 1 \times 10^{-3} \end{pmatrix}$ and $\vec{\vec{\alpha}}_2 = \begin{pmatrix} 4 \times 10^{-4} & 2 \times 10^{-4} & 2 \times 10^{-4} \\ 2 \times 10^{-4} & 4 \times 10^{-4} & 2 \times 10^{-4} \\ 2 \times 10^{-4} & 2 \times 10^{-4} & 1 \times 10^{-3} \end{pmatrix}$ which mainly differ in their off-diagonal elements, to compare with the classical LLG equation. We find that, since all components of the damping coefficients in the three cases are small, the differences among the various forms of damping coefficients are correspondingly insignificant. When the zz-component of the Gilbert damping tensor is big and the off-diagonal components are nonzero, the damping rate of z-component is slightly reduced while large off-diagonal components can partially compensate this reduction. In the long-time limit, the damping behaviors associated with different tensor forms converge and become indistinguishable, as shown in Fig.2. Although the quantitative differences from the classical LLG equation are relatively small for the parameters considered here, the present formulation provides a microscopic connection between spin-phonon interactions and the tensorial structure of Gilbert damping. Furthermore, compared with a scalar Gilbert damping coefficient, the tensorial damping struc- ture can capture direction-dependent energy dissipation and may become important in materials with stronger anisotropy or more complex lattice structure.

*Conclusion*—In this paper, we investigate the interactions that can give rise to the magnetic damping in two types of systems distinguished by their electronic transport properties, namely ferromagnetic conductor and magnetic insulator. By considering the specific interactions responsible for dissipation, we derive the corresponding interaction-induced damping terms within the framework of the LLG dynamics.

For ferromagnetic conductors, the s-d exchange interaction between itinerant electrons and localized magnetic moments leads to a damping contribution associated with spin accumulation and spin-transfer torque. The resulting damping tensor exhibits spatial dependence related to the variation of magnetization, reflecting the nonlocal characteristics of electron-mediated relaxation.

For magnetic insulators, we have considered spin-phonon coupling and derived the corresponding tensorial Gilbert damping from phonon angular momentum correlations. The anisotropic damping structure demonstrates that lattice degrees of freedom can generate damping characteristics different from those induced by electronic environments.

These results indicate that Gilbert damping is strongly determined by the microscopic nature of the coupled environment rather than being a universal scalar parameter. Different microscopic interactions, such as electron and phonon baths, can therefore lead to qualitatively different damping structures. The magnitude of the obtained damping coefficient for magnetic insulators is comparable to experimentally reported values, suggesting that spin-phonon coupling can provide a realistic contribution to magnetic relaxation.

The present framework focuses on the dominant damping mechanisms described by the considered interactions. In realistic materials, additional effects, including high-frequency phonon modes, other magnetic interactions, and crystal structures, may further modify the damping behavior. A systematic investigation of these effects and a quantitative determination of material parameters remain important directions for future studies. Nevertheless, our results provide a microscopic framework for understanding the relationship between environmental degrees of freedom and magnetic relaxation mechanisms.

*Acknowledgments*—We thank Prof. G. Su, Z.-G. Zhu, B. Gu, and Q.-B. Yan for their helpful discussions. This study is supported by the National Key R&D Program of China (Grant No.2022YFA1402703).

---

[1] V. G. Bar'yakhtar and B. A. Ivanov. The Landau-Lifshitz equation: 80 years of history, advances, and prospects. *Low Temperature Physics*, 41(9):663–669, 09 2015.

[2] T.L. Gilbert. A phenomenological theory of damping in ferromagnetic materials. *IEEE Transactions on Magnetics*, 40(6):3443–3449, 2004.

[3] Arseni Goussev, J. M. Robbins, Valeriy Slastikov, and Sergiy Vasylkevych. Dynamics of ferromagnetic domain walls under extreme fields. *Phys. Rev. B*, 101:020418, Jan 2020.

[4] Yuto Ohki and Masahito Mochizuki. Fundamental theory of current-induced motion of magnetic skyrmions. *J. Phys.: Condens. Matter*, 37(2):023003, oct 2024.

[5] Vladyslav M. Kuchkin, Pavel F. Bessarab, and Nikolai S. Kiselev. Thermal generation of droplet soliton in chiral magnet. *Phys. Rev. B*, 105:184403, May 2022.

[6] Olle Eriksson, Anders Bergman, Lars Bergqvist, and Johan Hellsvik. *Atomistic Spin Dynamics: Foundations and Applications*. Oxford University Press, 02 2017.

[7] R F L Evans, W J Fan, P Chureemart, T A Ostler, M O A Ellis, and R W Chantrell. Atomistic spin model simulations of magnetic nanomaterials. *J. Phys.: Condens. Matter*, 26(10):103202, feb 2014.

[8] Harry Suhl. *Relaxation Processes in Micromagnetics*. Oxford University Press, 06 2007.

[9] Ritwik Mondal, Marco Berritta, Ashis K. Nandy, and Peter M. Oppeneer. Relativistic theory of magnetic inertia in ultrafast spin dynamics. *Phys. Rev. B*, 96:024425, Jul 2017.

[10] Xiao Zhang, Zheng-Chuan Wang, Qing-Rong Zheng, Zheng-Gang Zhu, and Gang Su. Current-induced magnetic switching of a single molecule magnet on a spin valve. *Physics Letters A*, 379(9):848–852, 2015.

[11] Felipe Reyes-Osorio and Branislav K. Nikolić. Gilbert damping in metallic ferromagnets from schwinger-keldysh field theory: Intrinsically nonlocal, nonuniform, and made anisotropic by spin-orbit coupling. *Phys. Rev. B*, 109:024413, Jan 2024.

[12] Satadeep Bhattacharjee, Lars Nordstr$^{o}$m, and Jonas Fransson. Atomistic spin dynamic method with both damping and moment of inertia efects included from first principles. *Phys. Rev. Lett.*, 108:057204, Jan 2012.

[13] Mario Gaspar Quarenta, Mithuss Tharmalingam, Tim Ludwig, H. Y. Yuan, Lukasz Karwacki, Robin C. Verstraten, and Rembert A. Duine. Bath-induced spin inertia. *Phys. Rev. Lett.*, 133:136701, Sep 2024.

[14] Andreas Rückriegel and Peter Kopietz. Rayleigh-jeans condensation of pumped magnons in thin-film ferromagnets. *Phys. Rev. Lett.*, 115:157203, Oct 2015.

[15] L. Berger. Generation of dc voltages by a magnetic multilayer undergoing ferromagnetic resonance. *Phys. Rev. B*, 59:11465–11470, May 1999.

[16] Shufeng Zhang and Steven S.-L. Zhang. Generalization of the Landau-Lifshitz-Gilbert equation for conducting ferromagnets. *Phys. Rev. Lett.*, 102:086601, Feb 2009.

[17] Xiao-Tong Xu, Ya-Ru Wang, and Zheng-Chuan Wang. Stability analysis of skyrmionium driven by spin-polarized current. *Phys. Rev. B*, 111:054429, Feb 2025.

[18] N. Inaba, H. Asanuma, S. Igarashi, S. Mori, F. Kirino, K. Koike, and H. Morita. Damping constants of ni-fe and ni-co alloy thin films. *IEEE Transactions on Magnetics*, 42(10):2372–2374, 2006.

[19] Yuelei Zhao, Qi Song, See-Hun Yang, Tang Su, Wei Yuan, Stuart S. P. Parkin, Jing Shi, and Wei Han. Experimental investigation of temperature-dependent gilbert damping in permalloy thin films. *Sci. Rep.*, 6(1):22890, 2016.

[20] Mikihiko Oogane, Takeshi Wakitani, Satoshi Yakata, Resul Yilgin, Yasuo Ando, Akimasa Sakuma, and Terunobu Miyazaki. Magnetic damping in ferromagnetic thin films. *Japanese Journal of Applied Physics*, 45(5R):3889, may 2006.

[21] Ramesh Dhakal, Samuel Griffith, Kate Choi, and Stephen M. Winter. Spin-phonon coupling in transition metal insulators: General computational approach and application to $mnpse_3$, 2024.

[22] A. S. Ioselevich and H. Capellmann. Strongly correlated spin-phonon systems: A scenario for heavy fermions. *Phys. Rev. B*, 51:11446–11462, May 1995.

[23] Lifa Zhang and Qian Niu. Angular momentum of phonons and the einstein–de haas efect. *Phys. Rev. Lett.*, 112:085503, Feb 2014.

[24] Hailong Wang, Chunhui Du, P. Chris Hammel, and Fengyuan Yang. Spin transport in antiferromagnetic insulators mediated by magnetic correlations. *Phys. Rev. B*, 91:220410, Jun 2015.

[25] K. D. Belashchenko, O. Tchernyshyov, Alexey A. Kovalev, and O. A. Tretiakov. Magnetoelectric domain wall dynamics and its implications for magnetoelectric memory. *Applied Physics Letters*, 108(13):132403, 03 2016.